\documentclass[fleqn,10pt]{wlscirep}
\usepackage[utf8]{inputenc}
\usepackage{xcolor}
\usepackage{graphicx}
\usepackage{subfig}
\usepackage{cite}
\usepackage[font=small,labelfont=bf]{caption}

\title{Nanodiamond arrays on glass for quantification and fluorescence characterisation}

\author[1,*]{Ashleigh H. Heffernan}
\author[1]{Andrew D. Greentree}
\author[1]{Brant C. Gibson}
\affil[1]{ARC Centre of Excellence for Nanoscale BioPhotonics, School of Science, RMIT University, Melbourne, 3001, Australia}

\affil[*]{ashleigh.heffernan@rmit.edu.au}

\begin{abstract}
Quantifying the variation in emission properties of fluorescent nanodiamonds is important for developing their wide-ranging applicability. Directed self-assembly techniques show promise for positioning nanodiamonds precisely enabling such quantification. Here we show an approach for depositing nanodiamonds in pre-determined arrays which are used to gather statistical information about fluorescent lifetimes. The arrays were created via a layer of photoresist patterned with grids of apertures using electron beam lithography and then drop-cast with nanodiamonds. Electron microscopy revealed a 90\% average deposition yield across 3,376 populated array sites, with an average of 20 nanodiamonds per site. Confocal microscopy, optimised for nitrogen vacancy fluorescence collection, revealed a broad distribution of fluorescent lifetimes in agreement with literature. This method for statistically quantifying fluorescent nanoparticles provides a step towards fabrication of hybrid photonic devices for applications from quantum cryptography to sensing. 
\end{abstract}
\begin{document}

\flushbottom
\maketitle 
\thispagestyle{empty}
\noindent

\section*{Introduction}
Diamond has long been studied for its remarkable properties including chemical inertness, biocompatibility, transparency from the ultraviolet to the infra-red range, high thermal conductivity and mechanical strength \cite{Aharonovich2014, Wort2008, Mochalin2012, Krueger2008}. Recently, a large amount of research has focused on investigating the optical properties of fluorescent defects, often referred to as colour centres \cite{Aharonovich2014}. Detection of individual colour centres by confocal microscopy was demonstrated 20 years ago \cite{Gruber1997}, and over 500 active optical centres have so far been characterised \cite{Zaitsev2000}, many of which show single photon emission characteristics \cite{Aharonovich2011}.

One colour centre in particular has received the most attention: the nitrogen-vacancy (NV) centre, consisting of a substitutional nitrogen with a vacancy in its nearest neighbour lattice site \cite{Doherty2013}. The valence electrons of the nitrogen atom give rise to two possible charge configurations of this centre: neutral NV\textsuperscript{0}, or negatively charged NV\textsuperscript{-}. The emission spectrum of the NV\textsuperscript{-} centre is centred around 700 nm and its standard fluorescent lifetime in bulk diamond is $\sim$12 ns; both of these aspects make it attractive for biological imaging because autofluorescence from surrounding media can be avoided with spectral filters and\slash or time-gating \cite{Wu2013, Kuo2013}. Other aspects of the NV centre, including its long-lived and controllable spin state, make it attractive for quantum information purposes \cite{Grotz2012}. The NV centre has also been presented as a sensor, with changes in the environment (e.g. temperature \cite{Toyli2013}, magnetic field \cite{Balasubramanian2008}, or nearby nuclear spins \cite{Kost2015}) giving rise to variations in fluorescence, which is the basis of optically detected magnetic resonance (ODMR) measurements.

The excited state lifetime of the NV centre in diamond varies depending on several factors. The number of impurities in the diamond lattice, presence of non-diamond carbon material, size of the host crystal lattice (e.g. bulk single crystal diamond or nanodiamond), surface modification, and irradiation have all been reported to contribute to lifetime values ranging from 11.2 ns in the bulk to 25 ns \cite{Collins1983, Hanzawa1997, Beveratos2001, Tisler2009, Smith2010, Mona2013, McCloskey2014, Inam2013}.

Integrating nanodiamonds with mature photonic devices is essential to using their properties in a range of applications from quantum cryptography to sensing. Here we present a method for positioning as-received ball-milled nanodiamonds in pre-determined locations on glass, and use it as a tool for measuring the fluorescent lifetime statistics of the deposited material.

A number of approaches to locate NV centres in diamond have been developed over the last several years which fall into two broad categories: top-down direct-write techniques, and bottom-up assembly techniques.

Top-down approaches involve the creation of colour centres directly in bulk diamond using high-energy ion, neutron, or electron irradiation followed by annealing \cite{Mita1996, Wang2006, Orwa2011}. Implanting colour centres with nanoscale precision is a challenge, which can be overcome by using a lithographic mask \cite{Toyli2010, Spinicelli2011} or a direct-write focussed beam \cite{Huang2013, McCloskey2014}. Vertical distribution in the substrate is another challenge for direct writing techniques \cite{Santori2009, VlasovII2010, Rao2014}, especially since the depth of the colour centre below the surface has an impact on the fluorescence and to sense things near the surface the the colour centres have to be proximal to the surface.

Some bottom-up techniques involve nanodiamond particles, using them either as-received from the manufacturer or optically activating them by irradiating the nanodiamond powder before positioning \cite{Boudou2009}. The nanodiamonds can be deposited by spin-coating a solution on a substrate (leading to single particles dispersed randomly), pre-characterised, then positioned by microscopic probe (colloquially known as `pick and place’) \cite{vanderSar2009, Ampem-Lassen2009}; however, this method is time-consuming and not suited to large-scale production. 

An alternative to deterministic nanoparticle positioning is to use techniques of directed self-assembly.  They can be more efficient, have high spatial control of potential locations, and are often compatible with parallel fabrication methods \cite{Fulmes2015, Rivoire2009, Albrecht2013, Lee2010, Shimoni2014}. To optimise the particle surface properties for self-assembly, nanodiamonds often need significant chemical and physical processing. When suspended in solution, nanodiamonds acquire a zeta potential ($\sim$-30 mV as-received \cite{Shimoni2014}), which leads to readily formed agglomerates: to counteract this, the material can be treated with liquid phase purification or high temperature annealing \cite{Mochalin2012, Osswald2006}. Additional processes can be performed to enhance the fabrication, for example reactive ion etching has been used to attach a linker molecule before self-assembly \cite{Kianinia2016}. The method presented here keeps nanodiamond processing to a minimum, does not require any additional steps to process the substrate, and is a low-temperature fabrication method which is compatible with complementary metal-oxide-semiconductor (CMOS) processes.

Directed self-assembly suffers from the fact that pre-selected nanodiamonds cannot be placed at a given location, and the attachment specificity isn't always perfect. The practical harnessing of directed self-assembly therefore requires using defect-tolerant design considerations. If a device requires a certain number of elements to work, it may be more efficient to overcompensate during fabrication and have many defective elements than it is to laboriously fabricate every element perfectly and achieve 100 percent yield. In this context, self-assembly of nanodiamonds could lead to the creation of a large-scale many-qubit device \cite{Prawer2008}. Arrays such as the ones presented here are the first steps towards robust hybrid device fabrication on optically transparent substrates for quantum information processing, and towards novel large-area ODMR sensing and biological imaging.

\section*{Methods}
\subsection*{Array Fabrication}
A 3 inch [100] Si wafer was cleaned in an ultrasonic bath of acetone, then ethanol, then isopropyl alcohol (IPA), for 1 minute each. A 100 nm layer of polymethyl methacrylate (PMMA) A2 950 K was spin coated at 4000 rpm for 30 seconds, followed by a soft bake on a hot plate at 200$^{\circ}$C for 2 minutes. The wafer was then diced into $\sim$20 mm square samples. 

We used Si substrates to optimise the exposure process, and ultimately translated the technique to a transparent substrate: a glass slide (ProSciTech G300 white glass) was diced into $\sim$20 mm squares, and cleaned and spin coated in a similar method to the Si.

For the electron beam lithography (EBL) exposure process, an FEI Nova NanoSEM equipped with the Nabity Pattern Generation System software was used. The current of the 30kV electron beam was measured with a Faraday Cup; in high-vacuum mode ($ 1\times10^{-6}$ Torr for the silicon substrates) the current was measured to be 300 pA. The dwell time of the electron beam at each point was calculated by the NPGS software (using the measured value of the current). To mitigate the issue of electrical charge gathering on the surface of the glass slide samples, we used a low-vacuum mode in the SEM which introduces water vapour into the vacuum chamber (known as Variable Pressure Electron Beam Lithography, VP-EBL \cite{Myers2006}), at a pressure of 5 Torr and a measured beam current of 101 pA.

The pattern was designed to take into account the diffraction-limited resolution of confocal microscopy which was to be utilised post fabrication: to be confocally resolvable, the array sites needed to be at least 300 nm apart. We designed apertures ranging from 50 nm to 1 $\mu$m in diameter, in arrays with pitches ranging from 1.5 $\mu$m to 10 $\mu$m, and repeated the arrays over areas up to 15 mm$^2$.

The exposed patterns were developed in a 1:3 MIBK:IPA bath for 1 minute, then rinsed with IPA followed by water, and dried with pressurised air. To characterise the developed apertures without exposing the remaining photoresist, an atomic force microscope was used.

As-received nanodiamonds with a nominal diameter of 45 nm (Nabond Technologies, China) were suspended in Milli-Q water at a concentration of 1 $\mu$g/mL and sonicated for 10 minutes to break up the larger agglomerations. The zeta potential and average size of the remaining aggregates was measured with a Malvern Zetasizer, found to be -40 mV and and 124 nm respectively. The solution was drop cast on the developed photoresist and allowed to air dry overnight.

An acetone lift-off step was performed to strip the PMMA and top layer of nanodiamond material away, leaving behind nanodiamonds that had come into contact with the apertures. To remove remaining organic material (solvent or photoresist), the samples were placed in a plasma cleaner (Gatan Model 950, 65 W 13.56 MHz power supply) and subject to 2 minutes of H$_{2}$/O$_{2}$ plasma. Characterisation of the topography of the arrays was performed with an FEI Verios SEM, under low accelerating voltage and high stage bias without conductive coating.

\subsection*{Confocal Microscopy}
A custom confocal microscope \cite{Reineck2017} was used to characterise the NV\textsuperscript{-} fluorescence of the nanodiamond arrays at room temperature. A 532 nm continuous wave laser (LaserQuantum GEM 532) was used to excite the sample with 0.5 mW of excitation power. The beam was passed through a 532 nm line filter to filter out fluorescence from a single-mode fibre. A 561 nm dichroic mirror reflected the beam onto the back of a 100x objective with NA = 0.9. The same objective collected the fluorescence, which was passed through a 532 nm notch filter to filter out the reflected laser light. A 697 $\pm$ 75 nm band pass filter was used to block background light and transmit a region that would show the zero phonon lines and phonon sideband of the NV centre emission spectrum. The collected signal was split 90:10, with 10\% going to a spectrometer (Princeton Instruments SpectraPro 2500i with PIXIS 100BR camera) to confirm that the emission spectrum matched that of the NV centre, and the remaining 90\% being split 50:50 between two avalanche photodiode detectors.

For fluorescent lifetime measurements, a Fianium supercontinuum source was used, set to deliver 0.1 mW of 520 $\pm$ 10 nm light at a pulse rate of 20 MHz (corresponding to a repetition rate of 50 ns). The excitation and emission light was passed through the same filters as for the continuous wave excitation. The time-resolved direct fluorescence decay traces were obtained by a correlator card (Picoquant, TimeHarp 260).

We also excited the same sites with a 633 nm HeNe CW laser, and collected emission through a 633 nm notch filter and a 633 nm long pass filter, to look for silicon vacancy (SiV) centre emission in the as-received nanodiamond material.

\section*{Results and Discussion}

The fabrication of nanodiamond arrays using EBL has been published elsewhere, with a subsequent diamond film growth step \cite{Shimoni2014}. In this study, we used the same array fabrication method, but we present quantification of the deposition of nanodiamonds on a glass substrate, and we present fluorescent characterisation of the nanodiamond arrays without any other material processing.

Deformation of EBL patterns can be caused by electrical charge gathering on insulating substrates and deflecting the electron beam, however we found that the overall array pattern was not deformed, even on the glass substrate. We attribute the lack of deformation of our arrays to the use of VP-EBL and the simplicity of the overall pattern, each of which minimises the impact of charging. Charging can also be reduced by minimising the area of each aperture, but this requires a precisely focused electron beam. 200 nm circular apertures were written with the VP-EBL using a raster pattern for the exposure. The focus and alignment of the electron beam are critical requirements for accurate EBL, but attaining high precision of these in VP-EBL is challenging. The water vapour molecules that were introduced into the vacuum chamber to minimise charging can cause scattering in the electron beam and decrease the signal-to-noise ratio of the SEM image. Since the clarity of the image is essential to fine tuning the focus of the beam, a noisy image means a precisely focused beam is difficult to achieve.

\begin{figure*}[ht]
\centering
\includegraphics[width=1\linewidth]{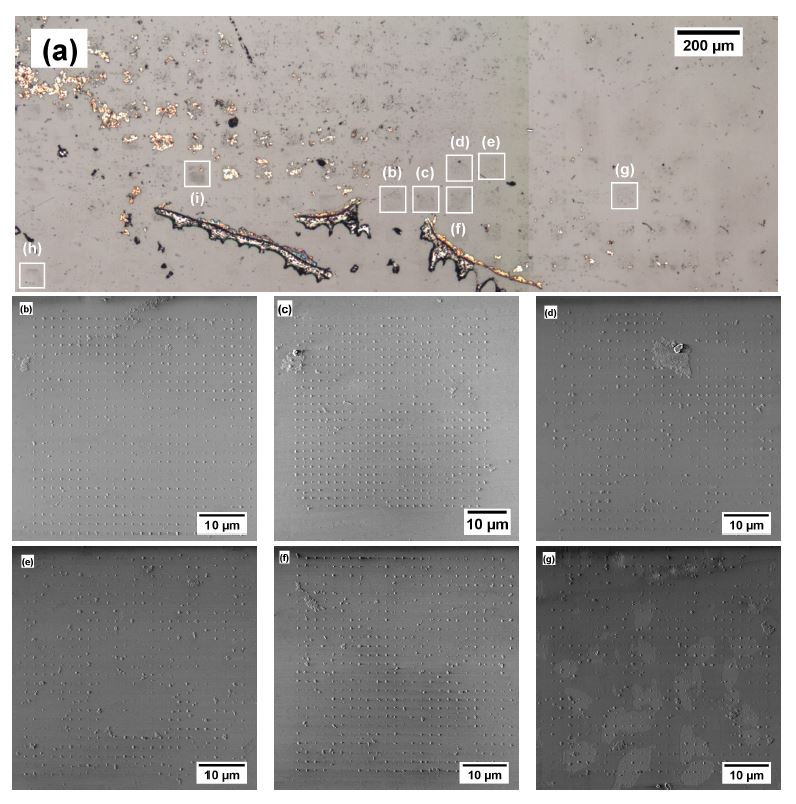}\\
\caption{Composite optical micrograph (a) and individual SEM micrographs (b)-(g) of nanodiamond arrays on a glass substrate. The pitch from array to array is 50 $\mu$m. Arrays (h) and (i) were studied by SEM and confocal microscope. Arrays (b)-(g) were only imaged by SEM. Each array has 623 sites (25 $ \times $ 25 with 2 on the lower right corner omitted for orientation purposes) and a pitch of 1.5 $\mu$m. The degree to which the self-assembly was successful has been quantified by eye and is shown in Table \ref{table:arraycoverage}.}
\label{fig:arraycoverage}
\end{figure*}

The deposition process left nanodiamond residue on the PMMA layer in typical coffee-ring patterns as the water evaporated and the droplet shrank. After stripping the photoresist, arrays of nanodiamond material were visible by optical microscope, as were some areas of non-specific deposition, as shown in Figure \ref{fig:arraycoverage}.

Two theories are proposed based on previous literature \cite{Shimoni2014} as to why the nanodiamonds remain at the aperture locations during the photoresist removal step. The negative zeta potential of the particles in solution may be responsible for a van der Waal attraction to the positively charged native silicon oxide layer on the wafer. Alternatively, the surface chemistry of the nanodiamonds (e.g. carboxylic groups) may result in covalent bonding with straggling polymer chains at the edges of the developed apertures in the photoresist. We often saw nanodiamonds in a circular pattern at each site, which may be evidence of the latter hypothesis; however this does not explain non-specific deposition, where nanodiamonds are attached to the substrate in places other than those pre-determined by EBL.

To quantify the yield of deposited nanodiamond material, six individual arrays were imaged by SEM (Figure \ref{fig:arraycoverage}) and the number of populated sites, displayed in Table \ref{table:arraycoverage} for each selected array, was deduced by visually counting the number of unpopulated sites. One of the arrays quantified showed 98\% yield (613 out of 623 sites populated with nanodiamond material), and the average total yield was 90\% over the 6 arrays considered. However, Figure \ref{fig:arraycoverage} shows some variability and the extent of non-specific deposition.

\begin{table} [tbh]
\centering
\begin{tabular} {cccc}
\hline
Array ID & Empty Sites & Populated Sites&Percent Yield\\
\hline
b & 10 & 613 & 98\\
\hline
c & 30 & 593 & 95\\
\hline
d & 65 & 558 & 89\\
\hline
e & 109 & 514 & 82\\
\hline
f & 42 & 581 & 93\\
\hline
g & 106 & 517 & 82\\
\hline
\hline
total & 362 & 3376 & 90\\
\end{tabular}
\caption{Quantification of nanodiamond material coverage of six of the eight arrays indicated in Figure \ref{fig:arraycoverage}. Each array was fabricated with 623 apertures developed in PMMA, and after deposition of nanodiamond material and removal of the photoresist, the yield was quantified by eye from SEM micrographs.}
\label{table:arraycoverage}
\end{table}

\begin{figure*}[ht]
\centering
{\includegraphics[width=1\linewidth]{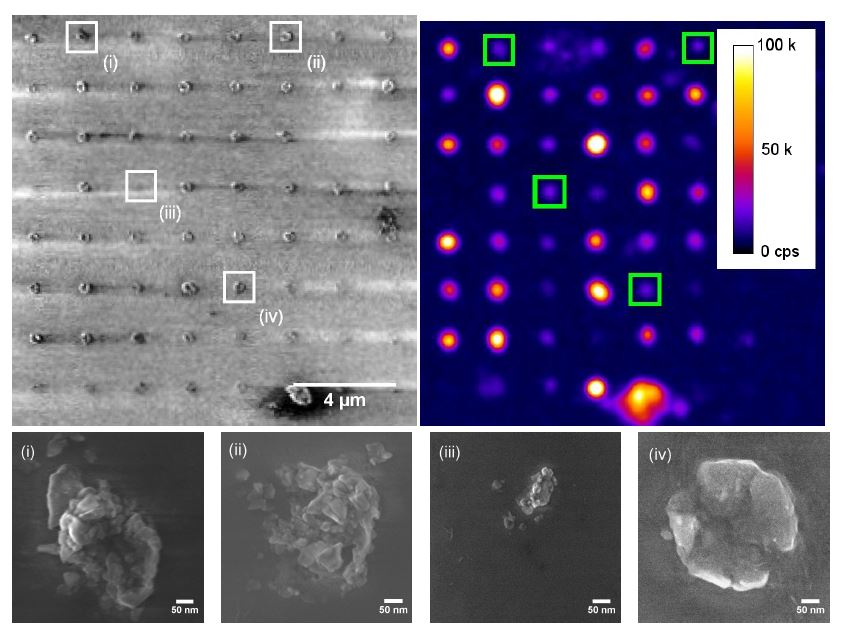}}
\caption{SEM (a) and confocal microscope (b) micrographs of one corner of array (h) from Figure \ref{fig:arraycoverage}. SEM micrographs of individual sites indicated (i, ii, iii, and iv) are shown below. The comparison of SEM to confocal micrographs highlights the variation of fluorescence from nanodiamond to nanodiamond: even though site (iii) has fewer particles than the other three, they all show similar fluorescence ($\sim$25 k counts per second).}
\label{fig:comparison6}
\end{figure*}

Nanodiamonds at individual sites were quantified from SEM micrographs for a selection of sites in arrays (h) and (i) shown in Figure \ref{fig:arraycoverage}. Close up SEM micrographs of several sites from array (h) are shown in Figure \ref{fig:comparison6}, displaying distributions of number and size of nanodiamonds per site.

Confocal microscopy was performed after developing and before drop-casting the nanodiamond material, and again after the photoresist was removed. The developed PMMA showed no outstanding fluorescence above background levels ($\sim$ 5000 counts per second).

Figure \ref{fig:comparison6} shows an SEM micrograph and a fluorescent map of a region of array (h) shown in Figure \ref{fig:arraycoverage}. Fluorescence spectra was collected from over 200 sites from arrays (h) and (i), for 30 seconds each. 153 of them displayed the characteristic NV centre broad phonon sideband (centred around 700 nm), although the zero phonon lines (NV \textsuperscript{0} at 575 nm and NV \textsuperscript{-} at 637 nm) were almost always too weak to distinguish. We also noted no obvious shifting of the phonon sideband. 

The quantification of nanodiamonds that do host NV centres in contrast to those that do not could be achieved with an array that has individual particles at each site. As the arrays hold multiple particles per site, we restrict our statistical analysis not to the number of nanodiamonds that host NV centres, but the number of sites that host at least one NV centre, and the number of nanodiamonds over 30 nm present (estimated visually from SEM micrographs). We note that the sites with NV fluorescence and those without show similar numbers of particles per site. The average for the former is 21 with a standard deviation of 10, and the average for the latter is 17 with a standard deviation of 9.

\begin{figure*}[ht]
\centering
{\includegraphics[width=1\linewidth]{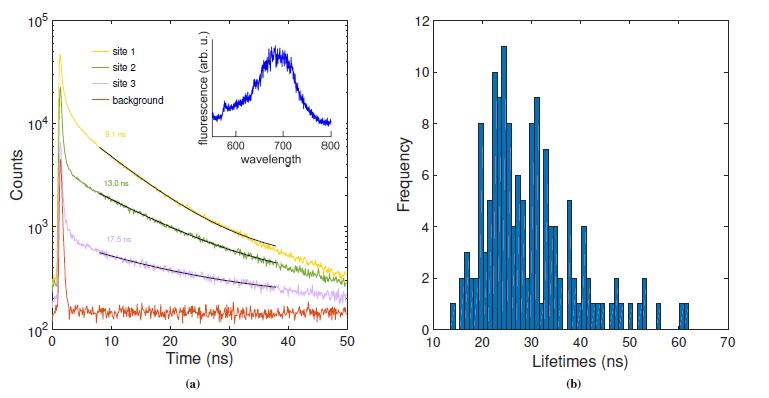}}
\caption{(a) Fluorescent decay traces from three different sites and a background measurement. The black lines are single exponential fits (presented for simplicity although multiple exponential fitting analysis was performed), and the values are the calculated average lifetimes of all the NV centre emission at that particular site. The inset is an example of a spectrum acquired, used to distinguish between sites that held NV centres and those that did not. (b) A histogram of the lifetimes obtained from the fluorescence time series fits.  The data set contained 153 fluorescent sites from a total of 220 analysed sites, and the lifetimes were calculated with a sum of weighted exponentials. We discarded lifetimes close to 5~ns and 100~ns as being artefacts of the fitting routine. The modal fluorescence lifetime was 24.5 ns, with a strongly asymmetric distribution towards longer lifetimes.  As expected, all of the measured lifetimes were greater than the spontaneous emission lifetime of NV centres in bulk diamond.}

\label{fig:analysis}
\end{figure*}

To collect time-dependent fluorescence data, we used pulse laser excitation on those sites that displayed typical NV fluorescence (153 sites in total from 220 sites from arrays (h) and (i) shown in Figure \ref{fig:arraycoverage}). Measuring the fluorescent intensity $I$ as a function of time $t$ allowed us to calculate the fluorescent lifetime $\tau$ of a site by fitting the data (with the Matlab routine lsqcurvefit) to the basic exponential decay equation $ I = \alpha  e^{-t/\tau} + \beta $ (with $\alpha$ and $\beta$ as scaling parameters to account for total fluorescence and background). Three representative lifetime traces are shown in Figure \ref{fig:analysis} (a), along with a measurement taken at an off-array site to gauge the instrument response and background fluorescence of the substrate.

Single exponential fitting gives a limited interpretation of the data, so more rigorous analysis was performed with a fitting function of a simple sum of weighted exponentials of the form $a + \sum_{i = 1}^{n} b_i \exp\left(-t/c_i\right)$ for $n$ exponentials with weighting $b_i$ and lifetime $c_i$, and background $a$.  Each time series was analysed for $n = 1, 2, 3, 4, 5$, and the best fit determined by choosing the number of exponential terms that maximised the adjusted R$^2$ parameter for the series.  As constraints, the first exponential was constrained $0 <c_1 100$~ns, and the other exponentials  constrained $5 < c_i < 100$~ns $i >1$. We discarded lifetimes close to 5~ns and 100~ns as being artefacts of the fitting routine.

Figure \ref{fig:analysis} (b) shows a histogram of the lifetimes obtained from the fluorescence time series fits.  The data set contained 153 fluorescent sites from a total of 220 analysed sites.  The modal fluorescence lifetime was 24.5 ns, with a strongly asymmetric distribution towards longer lifetimes.  As expected, all of the measured lifetimes were greater than the spontaneous emission lifetime of NV centres in bulk diamond.  Interestingly, most sites showed a single lifetime, and only two sites showed two distinct lifetimes in the range expected for NV centres (between 12 and 50 nm).  We do not interpret this as implying that most of the sites hosted single, isolated NV centres.  Instead it is likely that NV centres within a single site see similar refractive index environments, and hence radiative density of states, and therefore exhibit similar lifetimes.  Further analysis is required to fully understand the significance of this result. Fluorescent lifetime imaging experiments have shown similar lifetime distributions \cite{Tisler2009, GattoMonticone2013}, but the work presented here is the first time that this extent of statistical analysis has been conducted by direct measurement of NV centre lifetimes in as-received nanodiamonds.

Finally, we also performed a study of whether any SiV centres were present in the as received nanodiamond material. SiV centres are usually studied only after nanodiamond material has been irradiated \cite{Wang2006}, or after diamond films are grown via CVD under specific conditions to incorporate silicon atoms \cite{Zaitsev2001}. We observed no distinctive peaks at 737 nm corresponding to the zero phonon line of the SiV centre, which is unsurprising. Since the nanodiamonds used in this study were unprocessed, it is unlikely that they hosted many, if any, SiV centres.

\section* {Conclusion}

We have demonstrated a directed self-assembly method to position nanodiamonds in pre-determined locations at room temperature with minimal material processing. Despite an average of 10\% random assembly, the process is faster than manipulating individual particles, and the agility of EBL array pre-determination provides excellent spatial control. 

To advance the technique to a point where device fabrication is viable, particle attachment must be studied further. Recently published work \cite{Kianinia2016} suggests an optimal nanodiamond solution concentration to minimise agglomeration. Other researchers  \cite{Yoshikawa2015} have demonstrated ways to prevent agglomeration of nanodiamonds in solution and control their zeta potential, but they add degrees of complexity such as chemical processing and additive surface modification.

The ordered nature of arrays allows for identification of the same particles over different stages of an experiment. Potential future directions include characterising a set of nanodiamonds, deterministically changing their environment or modifying their surface, and re-characterising, with the aim of quantifying the modification.

Arrays of NV centres in bulk diamond have previously been used for magnetic imaging \cite{Steinert2013}, and single nanodiamonds have been used for single-molecule nuclear magnetic resonance \cite{Kost2015}. Our nanodiamond arrays provide initial steps for combining these two techniques for chemical sensing and imaging of biological systems, for example the surface of cells.

The work shown here also opens avenues for fabricating defect-tolerant hybrid photonic devices, consisting of, for example, nanodiamonds on dielectric substrates or waveguides. 

\bibliography{library}

\begin{thebibliography}{10}
\expandafter\ifx\csname url\endcsname\relax
  \def\url#1{\texttt{#1}}\fi
\expandafter\ifx\csname urlprefix\endcsname\relax\def\urlprefix{URL }\fi
\expandafter\ifx\csname doiprefix\endcsname\relax\def\doiprefix{DOI }\fi
\providecommand{\bibinfo}[2]{#2}
\providecommand{\eprint}[2][]{\url{#2}}

\bibitem{Aharonovich2014}
\bibinfo{author}{Aharonovich, I.} \& \bibinfo{author}{Neu, E.}
\newblock \bibinfo{journal}{\bibinfo{title}{{Diamond Nanophotonics}}}.
\newblock {\emph{\JournalTitle{Advanced Optical Materials}}}
  \textbf{\bibinfo{volume}{2}}, \bibinfo{pages}{911 -- 928}
  (\bibinfo{year}{2014}).
\newblock \doiprefix 10.1002/adom.201400189.

\bibitem{Wort2008}
\bibinfo{author}{Wort, C. J.~H.} \& \bibinfo{author}{Balmer, R.~S.}
\newblock \bibinfo{journal}{\bibinfo{title}{{Diamond as an electronic
  material}}}.
\newblock {\emph{\JournalTitle{Materials Today}}}
  \textbf{\bibinfo{volume}{11}}, \bibinfo{pages}{22--28}
  (\bibinfo{year}{2008}).
\newblock
  \urlprefix\url{http://www.sciencedirect.com/science/article/pii/S1369702107703498}.
\newblock \doiprefix http://dx.doi.org/10.1016/S1369-7021(07)70349-8.

\bibitem{Mochalin2012}
\bibinfo{author}{Mochalin, V.~N.}, \bibinfo{author}{Shenderova, O.},
  \bibinfo{author}{Ho, D.} \& \bibinfo{author}{Gogotsi, Y.}
\newblock \bibinfo{journal}{\bibinfo{title}{{The properties and applications of
  nanodiamonds}}}.
\newblock {\emph{\JournalTitle{Nat Nanotechnol}}} \textbf{\bibinfo{volume}{7}},
  \bibinfo{pages}{11--23} (\bibinfo{year}{2012}).
\newblock \urlprefix\url{http://www.ncbi.nlm.nih.gov/pubmed/22179567}.
\newblock \doiprefix 10.1038/nnano.2011.209.

\bibitem{Krueger2008}
\bibinfo{author}{Krueger, A.}
\newblock \bibinfo{journal}{\bibinfo{title}{{Diamond Nanoparticles: Jewels for
  Chemistry and Physics}}}.
\newblock {\emph{\JournalTitle{Advanced Materials}}}
  \textbf{\bibinfo{volume}{20}}, \bibinfo{pages}{2445--2449}
  (\bibinfo{year}{2008}).
\newblock \doiprefix 10.1002/adma.200701856.

\bibitem{Gruber1997}
\bibinfo{author}{Gruber, A.} \emph{et~al.}
\newblock \bibinfo{journal}{\bibinfo{title}{{Scanning Confocal Optical
  Microscopy and Magnetic Resonance on Single Defect Centers}}}.
\newblock {\emph{\JournalTitle{Science}}} \textbf{\bibinfo{volume}{276}},
  \bibinfo{pages}{2012--2014} (\bibinfo{year}{1997}).
\newblock \doiprefix 10.1126/science.276.5321.2012.

\bibitem{Zaitsev2000}
\bibinfo{author}{Zaitsev, A.~M.}
\newblock \bibinfo{journal}{\bibinfo{title}{{Vibronic spectra of
  impurity-related optical centers in diamond}}}.
\newblock {\emph{\JournalTitle{Physical Review B}}}
  \textbf{\bibinfo{volume}{61}}, \bibinfo{pages}{12909--12922}
  (\bibinfo{year}{2000}).
\newblock \urlprefix\url{http://link.aps.org/doi/10.1103/PhysRevB.61.12909}.
\newblock \doiprefix 10.1103/PhysRevB.61.12909.

\bibitem{Aharonovich2011}
\bibinfo{author}{Aharonovich, I.} \emph{et~al.}
\newblock \bibinfo{journal}{\bibinfo{title}{{Diamond-based single-photon
  emitters}}}.
\newblock {\emph{\JournalTitle{Reports on Progress in Physics}}}
  \textbf{\bibinfo{volume}{74}}, \bibinfo{pages}{76501} (\bibinfo{year}{2011}).
\newblock \doiprefix 10.1088/0034-4885/74/7/076501.

\bibitem{Doherty2013}
\bibinfo{author}{Doherty, M.~W.} \emph{et~al.}
\newblock \bibinfo{journal}{\bibinfo{title}{{The nitrogen-vacancy colour centre
  in diamond}}}.
\newblock {\emph{\JournalTitle{Physics Reports}}}
  \textbf{\bibinfo{volume}{528}}, \bibinfo{pages}{1--45}
  (\bibinfo{year}{2013}).
\newblock
  \urlprefix\url{http://www.sciencedirect.com/science/article/pii/S0370157313000562}.
\newblock \doiprefix 10.1016/j.physrep.2013.02.001.

\bibitem{Wu2013}
\bibinfo{author}{Wu, T.-J.} \emph{et~al.}
\newblock \bibinfo{journal}{\bibinfo{title}{{Tracking the engraftment and
  regenerative capabilities of transplanted lung stem cells using fluorescent
  nanodiamonds}}}.
\newblock {\emph{\JournalTitle{Nature Nanotechnology}}}
  \textbf{\bibinfo{volume}{8}}, \bibinfo{pages}{682--689}
  (\bibinfo{year}{2013}).
\newblock \urlprefix\url{http://dx.doi.org/10.1038/nnano.2013.147}.
\newblock \doiprefix 10.1038/nnano.2013.147.

\bibitem{Kuo2013}
\bibinfo{author}{Kuo, Y.}, \bibinfo{author}{Hsu, T.-Y.}, \bibinfo{author}{Wu,
  Y.-C.} \& \bibinfo{author}{Chang, H.-C.}
\newblock \bibinfo{journal}{\bibinfo{title}{{Fluorescent nanodiamond as a probe
  for the intercellular transport of proteins in vivo}}}.
\newblock {\emph{\JournalTitle{Biomaterials}}} \textbf{\bibinfo{volume}{34}},
  \bibinfo{pages}{8352--8360} (\bibinfo{year}{2013}).
\newblock \doiprefix 10.1016/j.biomaterials.2013.07.043.

\bibitem{Grotz2012}
\bibinfo{author}{Grotz, B.} \emph{et~al.}
\newblock \bibinfo{journal}{\bibinfo{title}{{Charge state manipulation of
  qubits in diamond.}}}
\newblock {\emph{\JournalTitle{Nature communications}}}
  \textbf{\bibinfo{volume}{3}}, \bibinfo{pages}{729} (\bibinfo{year}{2012}).
\newblock \urlprefix\url{http://dx.doi.org/10.1038/ncomms1729}.
\newblock \doiprefix 10.1038/ncomms1729.

\bibitem{Toyli2013}
\bibinfo{author}{Toyli, D.~M.} \emph{et~al.}
\newblock \bibinfo{journal}{\bibinfo{title}{{Fluorescence thermometry enhanced
  by the quantum coherence of single spins in diamond.}}}
\newblock {\emph{\JournalTitle{Proceedings of the National Academy of Sciences
  of the United States of America}}} \textbf{\bibinfo{volume}{110}}
  (\bibinfo{year}{2013}).
\newblock \doiprefix 10.1073/pnas.1306825110.
\newblock \eprint{1303.6730}.

\bibitem{Balasubramanian2008}
\bibinfo{author}{Balasubramanian, G.} \emph{et~al.}
\newblock \bibinfo{journal}{\bibinfo{title}{{Nanoscale imaging magnetometry
  with diamond spins under ambient conditions.}}}
\newblock {\emph{\JournalTitle{Nature}}} \textbf{\bibinfo{volume}{455}},
  \bibinfo{pages}{648--51} (\bibinfo{year}{2008}).
\newblock \urlprefix\url{http://dx.doi.org/10.1038/nature07278}.
\newblock \doiprefix 10.1038/nature07278.

\bibitem{Kost2015}
\bibinfo{author}{Kost, M.}, \bibinfo{author}{Cai, J.} \&
  \bibinfo{author}{Plenio, M.~B.}
\newblock \bibinfo{journal}{\bibinfo{title}{{Resolving single molecule
  structures with Nitrogen-vacancy centers in diamond.}}}
\newblock {\emph{\JournalTitle{Scientific reports}}}
  \textbf{\bibinfo{volume}{5}}, \bibinfo{pages}{11007} (\bibinfo{year}{2015}).
\newblock
  \urlprefix\url{http://www.nature.com/srep/2015/150605/srep11007/full/srep11007.html}.
\newblock \doiprefix 10.1038/srep11007.

\bibitem{Collins1983}
\bibinfo{author}{Collins, A.~T.}, \bibinfo{author}{Thomaz, M.~F.} \&
  \bibinfo{author}{Jorge, M. I.~B.}
\newblock \bibinfo{journal}{\bibinfo{title}{{Luminescence decay time of the
  1.945 eV centre in type Ib diamond}}}.
\newblock {\emph{\JournalTitle{Journal of Physics C: Solid State Physics}}}
  \textbf{\bibinfo{volume}{16}}, \bibinfo{pages}{2177--2181}
  (\bibinfo{year}{1983}).
\newblock
  \urlprefix\url{http://stacks.iop.org/0022-3719/16/i=11/a=020?key=crossref.e53541c5b1fb9e60387d4bdc3eaa68a1}.
\newblock \doiprefix 10.1088/0022-3719/16/11/020.

\bibitem{Hanzawa1997}
\bibinfo{author}{Hanzawa, H.}, \bibinfo{author}{Nisida, Y.} \&
  \bibinfo{author}{Kato, T.}
\newblock \bibinfo{journal}{\bibinfo{title}{{Measurement of decay time for the
  NV centre in Ib diamond with a picosecond laser pulse}}}.
\newblock {\emph{\JournalTitle{Diamond and Related Materials}}}
  \textbf{\bibinfo{volume}{6}}, \bibinfo{pages}{1595--1598}
  (\bibinfo{year}{1997}).
\newblock
  \urlprefix\url{http://linkinghub.elsevier.com/retrieve/pii/S092596359700037X}.
\newblock \doiprefix 10.1016/S0925-9635(97)00037-X.

\bibitem{Beveratos2001}
\bibinfo{author}{Beveratos, A.}, \bibinfo{author}{Brouri, R.},
  \bibinfo{author}{Gacoin, T.}, \bibinfo{author}{Poizat, J.-P.} \&
  \bibinfo{author}{Grangier, P.}
\newblock \bibinfo{journal}{\bibinfo{title}{{Nonclassical radiation from
  diamond nanocrystals}}}.
\newblock {\emph{\JournalTitle{Physical Review A}}}
  \textbf{\bibinfo{volume}{64}}, \bibinfo{pages}{061802}
  (\bibinfo{year}{2001}).
\newblock
  \urlprefix\url{http://journals.aps.org/pra/abstract/10.1103/PhysRevA.64.061802}.
\newblock \doiprefix 10.1103/PhysRevA.64.061802.

\bibitem{Tisler2009}
\bibinfo{author}{Tisler, J.} \emph{et~al.}
\newblock \bibinfo{journal}{\bibinfo{title}{{Fluorescence and Spin Properties
  of Defects in Single Digit Nanodiamonds}}}.
\newblock {\emph{\JournalTitle{ACS Nano}}} \textbf{\bibinfo{volume}{3}},
  \bibinfo{pages}{1959--1965} (\bibinfo{year}{2009}).
\newblock \urlprefix\url{http://dx.doi.org/10.1021/nn9003617}.
\newblock \doiprefix 10.1021/nn9003617.

\bibitem{Smith2010}
\bibinfo{author}{Smith, B.~R.}, \bibinfo{author}{Gruber, D.} \&
  \bibinfo{author}{Plakhotnik, T.}
\newblock \bibinfo{journal}{\bibinfo{title}{{The effects of surface oxidation
  on luminescence of nano diamonds}}}.
\newblock {\emph{\JournalTitle{Diamond and Related Materials}}}
  \textbf{\bibinfo{volume}{19}}, \bibinfo{pages}{314--318}
  (\bibinfo{year}{2010}).
\newblock \doiprefix 10.1016/j.diamond.2009.12.009.

\bibitem{Mona2013}
\bibinfo{author}{Mona, J.} \emph{et~al.}
\newblock \bibinfo{journal}{\bibinfo{title}{{Tailoring of structure, surface,
  and luminescence properties of nanodiamonds using rapid oxidative
  treatment}}}.
\newblock {\emph{\JournalTitle{Journal of Applied Physics}}}
  \textbf{\bibinfo{volume}{113}}, \bibinfo{pages}{114907}
  (\bibinfo{year}{2013}).
\newblock
  \urlprefix\url{http://scitation.aip.org/content/aip/journal/jap/113/11/10.1063/1.4795605}.
\newblock \doiprefix 10.1063/1.4795605.

\bibitem{McCloskey2014}
\bibinfo{author}{McCloskey, D.} \emph{et~al.}
\newblock \bibinfo{journal}{\bibinfo{title}{{Helium ion microscope generated
  nitrogen-vacancy centres in type Ib diamond}}}.
\newblock {\emph{\JournalTitle{Applied Physics Letters}}}
  \textbf{\bibinfo{volume}{104}}, \bibinfo{pages}{031109}
  (\bibinfo{year}{2014}).
\newblock
  \urlprefix\url{http://scitation.aip.org/content/aip/journal/apl/104/3/10.1063/1.4862331}.
\newblock \doiprefix 10.1063/1.4862331.

\bibitem{Inam2013}
\bibinfo{author}{Inam, F.~A.} \emph{et~al.}
\newblock \bibinfo{journal}{\bibinfo{title}{{Emission and Nonradiative Decay of
  Nanodiamond NV Centers in a Low Refractive Index Environment}}}.
\newblock {\emph{\JournalTitle{ACS Nano}}} \textbf{\bibinfo{volume}{7}},
  \bibinfo{pages}{3833--3843} (\bibinfo{year}{2013}).
\newblock \urlprefix\url{http://pubs.acs.org/doi/abs/10.1021/nn304202g}.
\newblock \doiprefix 10.1021/nn304202g.

\bibitem{Mita1996}
\bibinfo{author}{Mita, Y.}
\newblock \bibinfo{journal}{\bibinfo{title}{{Change of absorption spectra in
  type-Ib diamond with heavy neutron irradiation}}}.
\newblock {\emph{\JournalTitle{Physical Review B}}}
  \textbf{\bibinfo{volume}{53}}, \bibinfo{pages}{11360--11364}
  (\bibinfo{year}{1996}).
\newblock \urlprefix\url{http://link.aps.org/doi/10.1103/PhysRevB.53.11360}.
\newblock \doiprefix 10.1103/PhysRevB.53.11360.

\bibitem{Wang2006}
\bibinfo{author}{Wang, C.}, \bibinfo{author}{Kurtsiefer, C.},
  \bibinfo{author}{Weinfurter, H.} \& \bibinfo{author}{Burchard, B.}
\newblock \bibinfo{journal}{\bibinfo{title}{{Single photon emission from SiV
  centres in diamond produced by ion implantation}}}.
\newblock {\emph{\JournalTitle{Journal of Physics B: Atomic, Molecular and
  Optical Physics}}} \textbf{\bibinfo{volume}{39}}, \bibinfo{pages}{37--41}
  (\bibinfo{year}{2006}).
\newblock
  \urlprefix\url{http://stacks.iop.org/0953-4075/39/i=1/a=005?key=crossref.95838f4866b2b386f81cec0cae48c797}.
\newblock \doiprefix 10.1088/0953-4075/39/1/005.

\bibitem{Orwa2011}
\bibinfo{author}{Orwa, J.~O.} \emph{et~al.}
\newblock \bibinfo{journal}{\bibinfo{title}{{Engineering of nitrogen-vacancy
  color centers in high purity diamond by ion implantation and annealing}}}.
\newblock {\emph{\JournalTitle{Journal of Applied Physics}}}
  \textbf{\bibinfo{volume}{109}}, \bibinfo{pages}{083530}
  (\bibinfo{year}{2011}).
\newblock \urlprefix\url{http://aip.scitation.org/doi/10.1063/1.3573768}.
\newblock \doiprefix 10.1063/1.3573768.

\bibitem{Toyli2010}
\bibinfo{author}{Toyli, D.~M.}, \bibinfo{author}{Weis, C.~D.},
  \bibinfo{author}{Fuchs, G.~D.}, \bibinfo{author}{Schenkel, T.} \&
  \bibinfo{author}{Awschalom, D.~D.}
\newblock \bibinfo{journal}{\bibinfo{title}{{Chip-Scale Nanofabrication of
  Single Spins and Spin Arrays in Diamond}}}.
\newblock {\emph{\JournalTitle{Nano Letters}}} \textbf{\bibinfo{volume}{10}},
  \bibinfo{pages}{3168--3172} (\bibinfo{year}{2010}).
\newblock \urlprefix\url{http://dx.doi.org/10.1021/nl102066q}.
\newblock \doiprefix 10.1021/nl102066q.

\bibitem{Spinicelli2011}
\bibinfo{author}{Spinicelli, P.} \emph{et~al.}
\newblock \bibinfo{journal}{\bibinfo{title}{{Engineered arrays of
  nitrogen-vacancy color centers in diamond based on implantation of CN-
  molecules through nanoapertures}}}.
\newblock {\emph{\JournalTitle{New Journal of Physics}}}
  \textbf{\bibinfo{volume}{13}}, \bibinfo{pages}{025014}
  (\bibinfo{year}{2011}).
\newblock
  \urlprefix\url{http://iopscience.iop.org/article/10.1088/1367-2630/13/2/025014}.
\newblock \doiprefix 10.1088/1367-2630/13/2/025014.

\bibitem{Huang2013}
\bibinfo{author}{Huang, Z.} \emph{et~al.}
\newblock \bibinfo{journal}{\bibinfo{title}{{Diamond nitrogen-vacancy centers
  created by scanning focused helium ion beam and annealing}}}.
\newblock {\emph{\JournalTitle{Applied Physics Letters}}}
  \textbf{\bibinfo{volume}{103}}, \bibinfo{pages}{081906}
  (\bibinfo{year}{2013}).
\newblock
  \urlprefix\url{http://scitation.aip.org/content/aip/journal/apl/103/8/10.1063/1.4819339}.
\newblock \doiprefix 10.1063/1.4819339.

\bibitem{Santori2009}
\bibinfo{author}{Santori, C.}, \bibinfo{author}{Barclay, P.~E.},
  \bibinfo{author}{Fu, K.-M.~C.} \& \bibinfo{author}{Beausoleil, R.~G.}
\newblock \bibinfo{journal}{\bibinfo{title}{{Vertical distribution of
  nitrogen-vacancy centers in diamond formed by ion implantation and
  annealing}}}.
\newblock {\emph{\JournalTitle{Physical Review B}}}
  \textbf{\bibinfo{volume}{79}}, \bibinfo{pages}{125313}
  (\bibinfo{year}{2009}).
\newblock \urlprefix\url{http://link.aps.org/doi/10.1103/PhysRevB.79.125313}.

\bibitem{VlasovII2010}
\bibinfo{author}{Vlasov, I.} \emph{et~al.}
\newblock \bibinfo{journal}{\bibinfo{title}{{Nitrogen and luminescent
  nitrogen-vacancy defects in detonation nanodiamond}}}.
\newblock {\emph{\JournalTitle{Small}}} \textbf{\bibinfo{volume}{6}},
  \bibinfo{pages}{687--694} (\bibinfo{year}{2010}).
\newblock \urlprefix\url{http://www.ncbi.nlm.nih.gov/pubmed/20108229}.
\newblock \doiprefix 10.1002/smll.200901587.

\bibitem{Rao2014}
\bibinfo{author}{Rao, S.~G.} \emph{et~al.}
\newblock \bibinfo{journal}{\bibinfo{title}{{Directed assembly of nanodiamond
  nitrogen-vacancy centers on a chemically modified patterned surface.}}}
\newblock {\emph{\JournalTitle{ACS applied materials {\&} interfaces}}}
  \textbf{\bibinfo{volume}{6}}, \bibinfo{pages}{12893--900}
  (\bibinfo{year}{2014}).
\newblock \urlprefix\url{http://dx.doi.org/10.1021/am5027665}.
\newblock \doiprefix 10.1021/am5027665.

\bibitem{Boudou2009}
\bibinfo{author}{Boudou, J.-P.} \emph{et~al.}
\newblock \bibinfo{journal}{\bibinfo{title}{{High yield fabrication of
  fluorescent nanodiamonds.}}}
\newblock {\emph{\JournalTitle{Nanotechnology}}} \textbf{\bibinfo{volume}{20}},
  \bibinfo{pages}{235602} (\bibinfo{year}{2009}).
\newblock
  \urlprefix\url{http://www.pubmedcentral.nih.gov/articlerender.fcgi?artid=3201699{\&}tool=pmcentrez{\&}rendertype=abstract}.
\newblock \doiprefix 10.1088/0957-4484/20/23/235602.

\bibitem{vanderSar2009}
\bibinfo{author}{van~der Sar, T.} \emph{et~al.}
\newblock \bibinfo{journal}{\bibinfo{title}{{Nanopositioning of a diamond
  nanocrystal containing a single nitrogen-vacancy defect center}}}.
\newblock {\emph{\JournalTitle{Applied Physics Letters}}}
  \textbf{\bibinfo{volume}{94}}, \bibinfo{pages}{173104}
  (\bibinfo{year}{2009}).
\newblock
  \urlprefix\url{http://scitation.aip.org/content/aip/journal/apl/94/17/10.1063/1.3120558}.
\newblock \doiprefix 10.1063/1.3120558.

\bibitem{Ampem-Lassen2009}
\bibinfo{author}{Ampem-Lassen, E.} \emph{et~al.}
\newblock \bibinfo{journal}{\bibinfo{title}{{Nano-manipulation of diamond-based
  single photon sources}}}.
\newblock {\emph{\JournalTitle{Optics Express}}} \textbf{\bibinfo{volume}{17}},
  \bibinfo{pages}{11287} (\bibinfo{year}{2009}).
\newblock
  \urlprefix\url{http://www.osapublishing.org/viewmedia.cfm?uri=oe-17-14-11287{\&}seq=0{\&}html=true}.
\newblock \doiprefix 10.1364/OE.17.011287.

\bibitem{Fulmes2015}
\bibinfo{author}{Fulmes, J.} \emph{et~al.}
\newblock \bibinfo{journal}{\bibinfo{title}{{Self-aligned placement and
  detection of quantum dots on the tips of individual conical plasmonic
  nanostructures.}}}
\newblock {\emph{\JournalTitle{Nanoscale}}} \textbf{\bibinfo{volume}{7}},
  \bibinfo{pages}{14691--6} (\bibinfo{year}{2015}).
\newblock
  \urlprefix\url{http://pubs.rsc.org/en/content/articlehtml/2015/nr/c5nr03546e}.
\newblock \doiprefix 10.1039/c5nr03546e.

\bibitem{Rivoire2009}
\bibinfo{author}{Rivoire, K.} \emph{et~al.}
\newblock \bibinfo{journal}{\bibinfo{title}{{Lithographic positioning of
  fluorescent molecules on high-Q photonic crystal cavities}}}.
\newblock {\emph{\JournalTitle{Applied Physics Letters}}}
  \textbf{\bibinfo{volume}{95}}, \bibinfo{pages}{123113}
  (\bibinfo{year}{2009}).
\newblock
  \urlprefix\url{http://scitation.aip.org/content/aip/journal/apl/95/12/10.1063/1.3232233}.
\newblock \doiprefix 10.1063/1.3232233.

\bibitem{Albrecht2013}
\bibinfo{author}{Albrecht, A.} \emph{et~al.}
\newblock \bibinfo{journal}{\bibinfo{title}{{Self-assembling hybrid
  diamond–biological quantum devices}}}.
\newblock {\emph{\JournalTitle{New Journal of Physics}}}
  \textbf{\bibinfo{volume}{16}}, \bibinfo{pages}{093002}
  (\bibinfo{year}{2014}).
\newblock
  \urlprefix\url{http://stacks.iop.org/1367-2630/16/i=9/a=093002?key=crossref.a9845e3d4972d3b1b26cb6d900e179b6}.
\newblock \doiprefix 10.1088/1367-2630/16/9/093002.

\bibitem{Lee2010}
\bibinfo{author}{Lee, S.-K.}, \bibinfo{author}{Kim, J.-H.},
  \bibinfo{author}{Jeong, M.-G.}, \bibinfo{author}{Song, M.-J.} \&
  \bibinfo{author}{Lim, D.-S.}
\newblock \bibinfo{journal}{\bibinfo{title}{{Direct deposition of patterned
  nanocrystalline CVD diamond using an electrostatic self-assembly method with
  nanodiamond particles}}}.
\newblock {\emph{\JournalTitle{Nanotechnology}}} \textbf{\bibinfo{volume}{21}},
  \bibinfo{pages}{505302} (\bibinfo{year}{2010}).
\newblock \doiprefix 10.1088/0957-4484/21/50/505302.

\bibitem{Shimoni2014}
\bibinfo{author}{Shimoni, O.} \emph{et~al.}
\newblock \bibinfo{journal}{\bibinfo{title}{{Development of a Templated
  Approach to Fabricate Diamond Patterns on Various Substrates}}}.
\newblock {\emph{\JournalTitle{ACS Applied Materials {\&} Interfaces}}}
  \textbf{\bibinfo{volume}{6}}, \bibinfo{pages}{8894--8902}
  (\bibinfo{year}{2014}).
\newblock \doiprefix 10.1021/am5016556.

\bibitem{Osswald2006}
\bibinfo{author}{Osswald, S.}, \bibinfo{author}{Yushin, G.},
  \bibinfo{author}{Mochalin, V.}, \bibinfo{author}{Kucheyev, S.~O.} \&
  \bibinfo{author}{Gogotsi, Y.}
\newblock \bibinfo{journal}{\bibinfo{title}{{Control of sp2/sp3 carbon ratio
  and surface chemistry of nanodiamond powders by selective oxidation in
  air.}}}
\newblock {\emph{\JournalTitle{Journal of the American Chemical Society}}}
  \textbf{\bibinfo{volume}{128}}, \bibinfo{pages}{11635--42}
  (\bibinfo{year}{2006}).
\newblock \urlprefix\url{http://dx.doi.org/10.1021/ja063303n}.
\newblock \doiprefix 10.1021/ja063303n.

\bibitem{Kianinia2016}
\bibinfo{author}{Kianinia, M.} \emph{et~al.}
\newblock \bibinfo{journal}{\bibinfo{title}{{Robust, directed assembly of
  fluorescent nanodiamonds}}}.
\newblock {\emph{\JournalTitle{Nanoscale}}} \textbf{\bibinfo{volume}{8}},
  \bibinfo{pages}{18032--18037} (\bibinfo{year}{2016}).
\newblock \urlprefix\url{http://xlink.rsc.org/?DOI=C6NR05419F}.
\newblock \doiprefix 10.1039/C6NR05419F.

\bibitem{Prawer2008}
\bibinfo{author}{Prawer, S.} \& \bibinfo{author}{Greentree, A.~D.}
\newblock \bibinfo{journal}{\bibinfo{title}{{Diamond for Quantum Computing}}}.
\newblock {\emph{\JournalTitle{Science}}} \textbf{\bibinfo{volume}{320}}
  (\bibinfo{year}{2008}).

\bibitem{Myers2006}
\bibinfo{author}{Myers, B.~D.} \& \bibinfo{author}{Dravid, V.~P.}
\newblock \bibinfo{journal}{\bibinfo{title}{{Variable Pressure Electron Beam
  Lithography (VP- e BL): A New Tool for Direct Patterning of Nanometer-Scale
  Features on Substrates with Low Electrical Conductivity}}}.
\newblock {\emph{\JournalTitle{Nano Letters}}} \textbf{\bibinfo{volume}{6}},
  \bibinfo{pages}{963--968} (\bibinfo{year}{2006}).
\newblock \urlprefix\url{http://dx.doi.org/10.1021/nl0601278}.
\newblock \doiprefix 10.1021/nl0601278.

\bibitem{Reineck2017}
\bibinfo{author}{Reineck, P.} \emph{et~al.}
\newblock \bibinfo{journal}{\bibinfo{title}{{Bright and photostable
  nitrogen-vacancy fluorescence from unprocessed detonation nanodiamond}}}.
\newblock {\emph{\JournalTitle{Nanoscale}}} \textbf{\bibinfo{volume}{9}},
  \bibinfo{pages}{497--502} (\bibinfo{year}{2017}).
\newblock \urlprefix\url{http://xlink.rsc.org/?DOI=C6NR07834F}.
\newblock \doiprefix 10.1039/C6NR07834F.

\bibitem{GattoMonticone2013}
\bibinfo{author}{{Gatto Monticone}, D.} \emph{et~al.}
\newblock \bibinfo{journal}{\bibinfo{title}{{Systematic study of defect-related
  quenching of NV luminescence in diamond with time-correlated single-photon
  counting spectroscopy}}}.
\newblock {\emph{\JournalTitle{Physical Review B}}}
  \textbf{\bibinfo{volume}{88}}, \bibinfo{pages}{155201}
  (\bibinfo{year}{2013}).
\newblock \urlprefix\url{http://link.aps.org/doi/10.1103/PhysRevB.88.155201}.
\newblock \doiprefix 10.1103/PhysRevB.88.155201.

\bibitem{Zaitsev2001}
\bibinfo{author}{Zaitsev, A.~M.}
\newblock \emph{\bibinfo{title}{{Optical Properties of Diamond}}}
  (\bibinfo{publisher}{Springer Berlin Heidelberg}, \bibinfo{address}{Berlin,
  Heidelberg}, \bibinfo{year}{2001}).
\newblock \urlprefix\url{http://link.springer.com/10.1007/978-3-662-04548-0}.

\bibitem{Yoshikawa2015}
\bibinfo{author}{Yoshikawa, T.} \emph{et~al.}
\newblock \bibinfo{journal}{\bibinfo{title}{{Appropriate salt concentration of
  nanodiamond colloids for electrostatic self-assembly seeding of monosized
  individual diamond nanoparticles on silicon dioxide surfaces.}}}
\newblock {\emph{\JournalTitle{Langmuir : the ACS journal of surfaces and
  colloids}}} \textbf{\bibinfo{volume}{31}}, \bibinfo{pages}{5319--25}
  (\bibinfo{year}{2015}).
\newblock \urlprefix\url{http://dx.doi.org/10.1021/acs.langmuir.5b01060}.
\newblock \doiprefix 10.1021/acs.langmuir.5b01060.

\bibitem{Steinert2013}
\bibinfo{author}{Steinert, S.} \emph{et~al.}
\newblock \bibinfo{journal}{\bibinfo{title}{{Magnetic spin imaging under
  ambient conditions with sub-cellular resolution.}}}
\newblock {\emph{\JournalTitle{Nature communications}}}
  \textbf{\bibinfo{volume}{4}}, \bibinfo{pages}{1607} (\bibinfo{year}{2013}).
\newblock \urlprefix\url{http://dx.doi.org/10.1038/ncomms2588}.
\newblock \doiprefix 10.1038/ncomms2588.

\end{thebibliography}

\section*{Acknowledgements}
The authors acknowledge the facilities, and the scientific and technical assistance of the RMIT University's Microscopy \& Microanalysis Facility, a linked laboratory of the Australian Microscopy \& Microanalysis Research Facility. We also thank Dr. Desmond Lau for assistance with the confocal microscope studies. This work has been supported by ARC grants (FT110100225, FT160100357, LE140100131, CE140100003).

\section*{Author contributions statement}
A.H. conducted experiments and wrote the manuscript, B.G. and A.G. conceived the experiments and reviewed the manuscript. All authors analysed and interpreted the data.

\section*{Additional information}
\textbf{Competing financial interests:} The authors declare no competing financial interests.

\end{document}